\documentclass{iau}
\usepackage{graphicx}

\title[JD 11.~~SBS galaxies in seven fields] 
{Complex investigation of SBS galaxies\\ in seven selected fields}

\author[S.Hakopian]   
{S.A. Hakopian
 \ }

\affiliation{Byurakan Astrophysical Observatory, (BAO), Byurakan, 0213, Armenia.
\\email:  {\tt susanaha@bao.sci.am} \\[\affilskip] {}}

\pubyear{2008}
\volume{xxx}  
\pagerange{119--126}
\jname{Title of your IAU Symposium}
\editors{A.C. Editor, B.D. Editor \& C.E. Editor, eds.}
\usepackage{amsmath}

\begin{document}

\maketitle

\begin{abstract}

It is known that the main criterium for the selection of active objects in the First Byurakan, otherwise Markarian survey, was the presence of uv-excess on low-dispersion spectra registered on photographic plates obtained with the 1-m Shmidt type telescope of Byurakan in Armenia. Using the presence of emission lines as the second criterium became possible during the Second Byurakan Survey because of its improved technique. Through this criterium a majority of objects, extended by morphology (i.e. not stellated), were selected into the separate "sample of galaxies" of SBS. Certainly, there were cases of untrue selection, particularly, on faint magnitudes, when absorption lines were taken for emission ones and so on. With the program, briefly presented below, a study of SBS galaxies, including evaluation of  an effectivity of selection criteria, was undertaken by means of complex investigation of their very representative part, pooled in our basic sample. The completion of the follow-up slit spectroscopy of these about 500 objects of up to 19.5 apparent magnitude, followed by obtaining of the redshifts and absolute magnitudes, formed the main stage of implementation of the program. Observations were carried out with the 6-m telescope of SAO Russia and the 2.6-m Byurakan telescope. Also, the scheme, adapted to spectral material, was developed to provide homogeneous classification, directed, in particular, to separate galaxies of AGN activity, dividing them by well-known types, and starforming, SfG, activity. As a first step in classification of starforming galaxies, which constitute more than 80\%\ of the basic sample, we provided two classes, SfGcontinual and SfGnebular. Averaged statistics of our SfG galaxies show, that every fifth of them is in more active, nebular phase of starforming activity, most of which are known as blue compact galaxies. Especially in connection with this it must be noted, that, by the analysis, namely for the latter objects, the effectiveness of the survey is the highest, so that blue compact galaxies represent the best product of SBS among extended objects. Aimed on further specifications in classification of SfG galaxies, other generalizations and statistics in frames of ongoing investigation, detailed studies of individual galaxies are currently beeing held. In the base are data of panoramic spectroscopy obtained using  multipupil spectrographs MPFS and VAGR, on observations with the 6-m telescope of SAO Russia and the 2.6-m telescope of Byurakan, correspondently.  
\end{abstract}

\firstsection 
\section{Introduction}

The Second Byurakan Spectral Sky Survey (SBS) (\cite[Markaryan \& Stepanyan, 1983]{MarkaryanStepanyan83}) was undertaken by Markarian as 
a direct continuation of the First Byurakan Survey (FBS) (\cite[Markaryan, 1967]{Markaryan67}) to achieve magnitudes fainter than in FBS in search of active objects. Both surveys were carried out with the 1-m Shmidt type telescope of Byurakan observatory with the use of objective prisms of the same size as the entrance port of the telescope. Main differences of conducting the two surveys are seen on Table 1. Besides, it must be noted that the objects in SBS have been selected not only by presence of uv-excess on their low dispersion spectra, as it was done in FBS, but also by the presence of emission lines, as the second selection criterium. More than 1500 objects, which have been included in the common sample of FBS, are more famous as "Markarian galaxies". About 3000 objects selected in the 65 fields, each of 16sq. deg., which compile the SBS area, are distributed in the three samples, two of which (the sample of quasars and stars) consist of starlike objects.\\  
 In the third one, the sample of galaxies, initially about 1300 extended objects were included. Later, with the help of sources other than the SBS photographic plates, 200 more objects have been added to this list. As homogeneity of the object selection is important for the purposes of our program, we are working with the original list of SBS galaxies.

\begin{table}
  \begin{center}
  \caption{ Main differences of conducting the two byurakan surveys.}
  \label{tab1}
 {\scriptsize
  \begin{tabular}{|l|c|ccc|}\hline 
{\bf Survey} &  {\bf FBS  } & {\bf}& {\bf SBS } & {}\\\hline
{Limiting magnitudes (pg)} & $ {\sim 17.5} $ & &${\sim 19.5}$&\\ \hline
{Objective prism (angular degree)} & {1.5}  & $ {1.5} $&
 $  { 3} $& $   { 4}$ \\ \hline
Photographic plate &  Kodak IIa-F  & {Kodak IIIa-J  } &  {Kodak IIIa-J} & {Kodak IIIa-F } \\ \hline\
Hypersensitization of the plates  &- &{i n}&{n i t r o g e n}& {   v a  p o r }    \\ \hline
The used filter & - & {- } &  { GG495} & {RG2 } \\ \hline
 Exposure time for lim mag (min) &15 & {30-60}& {60-120}& {630-690}\\ \hline
Dispersion (\AA/mm)& 2500 (H\begin{tiny}$\beta$\end{tiny})  & 1800 (H\begin{tiny}$\beta$\end{tiny}) & 850 (H\begin{tiny}$\gamma$\end{tiny}) & 1100 (H\begin{tiny}$\alpha$\end{tiny})  \\ \hline
Spectral range (nm)& 350-690 & 350-540 & 490-540 & 630-690\\ \hline
Total area (square degrees)& $\sim1700$ & $\sim1000$ & $\sim160$ & $\sim800$\\ \hline
  \end{tabular}}
 \end{center}
\vspace{1mm}
 \scriptsize
\end{table}

\begin{figure}[]
\begin{center}
 \includegraphics[width=4.5in]{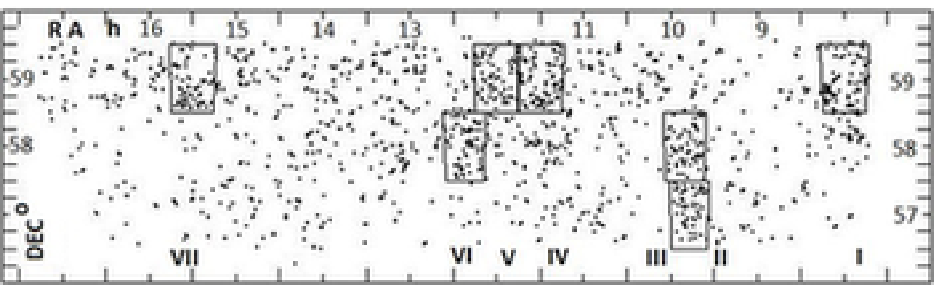} 
 \caption \tiny
 {Boundaries of selected fields within distribution of SBS galaxies in the whole area of SBS.}
   \label{fig1}
\end{center}
\end{figure}

\section{Accomplished Stages of the Program}
\textbf{2.1\textit{ Construction of  the basic sample}.}\\
More than a third part of the whole initial sample of SBS galaxies, about 500, constituted our basic sample. Namely, it is composed of those selected as galaxies in the seven of 65 in all fields of the survey (\cite[Hakopian \& Balayan 1999]{HakopianBalayan99}). The selected fields are numbered in order of ascending of right ascension of their centers, as shown on Fig.1, along with fields' boundaries given within the boundaries of the whole SBS area, in which the distribution of all SBS galaxies is depicted too. The second column of the table 2 gives initial numerical distribution of the galaxies by the seven fields (after excluding of the obviously wrong objects). These fields differ from many others, in particular by observations, completed with all the three objective prisms, high quality of the obtained photographic plates, providing the best for the survey  limiting magnitudes (19.0 -19.5) and so on. The level of completeness, calculated for the corresponding seven subsamples of SBS galaxies, according to the test V/Vmax (\cite[Shmidt 1968]
{Shmidt68}), was in the range 17.5 - 19.5 mag, i.e. higher than the value 16.5 - 17.0 mag, averaged for all 65 fields of the survey. The stated above is enough to claim, that our basic sample, being in addition statistically representative, must reflect productivity of the survey along the entire scale of apparent magnitudes, including the limiting ones (\cite [Hakopian 2002]{Hakopian02}).\\

\textbf{2.2. \textit{ Completion of the follow-up slit spectroscopy}.}\\
For the accomplishment of the follow-up slit spectroscopy of all the objects of the initial basic sample in frames of this program observations of more than 200 predominantly faint (17.5$^{\begin{tiny}m\end{tiny}}$ - 
19.5$^{\begin{tiny}m\end{tiny}}$) objects have been undertaken for the first time and observations of about 30 objects have been repeated (\cite [Hakopian 2002]{Hakopian02}, \cite [Hakopian \& Balayan 2002.1]{HakopianBalayan02.1} and references therein, \cite [Hakopian \& Balayan 2004]{HakopianBalayan04}). The observations were carried out from 1996 to 2002 with the 6-m telescope of Special Astrophysical Observatory of Russian Academy of Sciences (SAO RAS) (http://www.sao.ru), in combination with the Long Slit Spectrograph (LSS) in  different modifications and Multi Pupil Fiber Spectrograph (MPFS) and with the 2.6-m Byurakan telescope in Armenia in combination with the focal cameras ByuFOSC and SCORPIO in its spectral mode. The best values for dispersion of spectral material (1.3 \AA/pix) were obtained with the MPFS, the worst marks (5 \AA/pix), with the LSS in 1996.\\
  
\textbf{2.3  \emph{Development of the classification scheme.}}\\
On the basis of standard diagnostic criteria a scheme adapted to the spectral material has been developed to conduct preliminary and homogeneous classification of all objects of the basic sample.  The first purpose was to confirm the presence of signs of activity on the slit spectra, then, to identify AGN or starforming activity, using the ratio of forbidden and allowed lines 
(\cite[Veilleux \& Osterbrock 1987]{VeilleuxOsterbrock87}). Galaxies of starforming activity of any type (starburst, HII, BCG, etc.) have been combined under the designation SfG (Starforming Galaxy) with two subclasses, which are formed by analogy with the scheme of Terlevich (\cite [Terlevich 1997]{Terlevich97}). As seen from the structure of the scheme, shown on Fig.2, the AGN subclasses are  widely known, with the exception of LINER 1, the description of which is given  below, along with the description of the remaining ones.\\

\begin{figure}[h]
\begin{center}
 \includegraphics[width=4.5in]{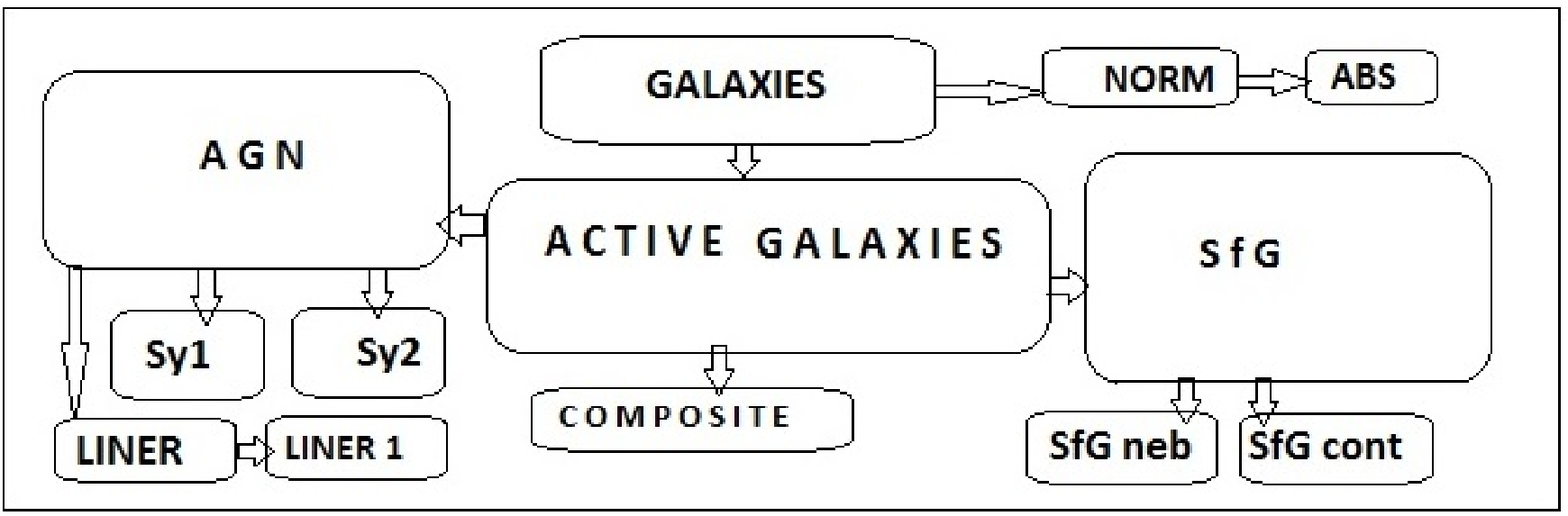}
 \caption{The structure of the adapted scheme used for objects' classification.}
   \label{fig2}
\end{center}
\end{figure}

\textbf{2.3.1.\emph{Description of classes in the used scheme.}}\\
\textbf{ AGN classes:}\\ \textbf{Sy1} -  galaxies of type Sy1.5 and lower ( \cite[Khachikyan \&Weedman 1974]{KhachikyanWeedman74}, \cite[Osterbrock 1993]{Osterbrock93});
\textbf{Sy2} - [NII]$\lambda$6583/H$\alpha$$>$0.7 and [OIII]$\lambda$5007/H$\alpha$$>$3; galaxies of type Sy1.6 and higher \\(\cite[Khachikyan \& Weedman 1974]{KhachikyanWeedman74},  \cite[Osterbrock 1993]{Osterbrock93});\\ \textbf{LINER} - correspondence  of at least two ratios of spectral lines intensities to the classical definition ( \cite[Heckman 1980]{Heckman80});\\ \textbf{LINER1} - a one dimensional LINER - [NII]$\lambda$6583/H$\alpha$$>$0.7 (by analogy with \cite [Filippenko \& Sargent 1985]{FilippenkoSargent85}), with or without data on the [OIII]$\lambda$5007 and H$\alpha$ lines;\\
\textbf{Composite} - galaxies with signs of both nuclear and starforming activity (by analogy with \cite[Veron et al. 1997]{Veron etal97}).\\
\textbf{SfG classes:}\\\textbf{ SfGneb} - galaxies in the nebular starforming phase, referred to as the nebular phase in Terlevich's scheme ( \cite [Terlevich 1997]{Terlevich97}) and distinguished, in particular, by having EW(H$\alpha$)$>$100;\\ \textbf{SfGcont} - galaxies in the continual starforming phase, referred to as one of the early continual or the later continual phases in Terlevich's scheme ( \cite [Terlevich 1997]{Terlevich97}), for which the equation EW(H$\alpha$)$<$100 stays common;\\
\textbf{Norm} - a galaxy in  emission spectrum of which only weak H$\alpha$ is present, EW(H$\alpha)$$<$5, with or without absorption lines (by analogy with \cite[Gioia et al. 1990]{Gioia_etal90});\\
\textbf{Abs} - a galaxy with spectrum including only absorption lines, when the spectral region comprising H$\alpha$ line is available 
(\cite[Bica \& Alloin 1987]{BicaAlloin87}).\\

\begin{table}[h]
  \begin{center}
  \caption{Common initial and final distribution of exploring galaxies by classes.}
  \label{tab2}
 {\scriptsize
  \begin{tabular}{|c||c||ccc|cc|c|c|}\hline 
{\bf Selected }&{\bf initial}&{\bf }& {\bf AGN}&{\bf}&{\bf SfG}&{}&{\bf Norm}&{\bf Abs}\\ 
{\bf Field }&{\bf number}&{\bf Sy1}&{\bf Sy2}&{\bf LINER}&{\bf SfGneb}&
{\bf SfGcont}&{\bf }&{\bf} \\ \hline
 {I}&{78}&{1}&{3}&{0}&{10}&{46} & {10}& {8}\\ \hline
 {II}&{52}&{3}&{1}&{0}&{4}&{28}&{5}&{11}\\ \hline
 {III}&{65}&{1}&{0}&{0}&{4}&{38}&{8}&{14}\\ \hline 
  {IV}&{90}&{5}&{2}&{0}&{17}&{37}&{19}&{10}\\ \hline
 {V}&{68}&{1}&{1}&{1}&{4}&{37}&{16}&{8}\\ \hline
 {VI}&{71}&{3}&{0}&{1}&{11}&{50}&{4}&{2}\\ \hline
 {VII}&{65}&{4}&{3}&{1}&{5}&{23}&{8}&{21}\\ \hline 
  {}&{}&{}&{}&{}&{}&{}&{}&{}\\  
 {sum}&{489}&{18}&{10}&{3}&{55}&{259}&{70}&{74}\\ \hline 
 \end{tabular}}
 \end{center}
\vspace{1mm}
 \scriptsize
\end{table}

\textbf{2.4.\emph {Final samples and some of analytical results.}}\\
Table 2 gives distribution of the objects of the basic sample in final subsamples, organized by classes. With the help of these data it is easy to find out that  the percentage of galaxies with confirmed activity is about 70 \%, relative to the initial number of objects in the basic sample, that the most of these, in their turn, are starforming galaxies and so on.\\
The detailed analysis of the distribution of apparent and absolute magnitudes and, certainly, of the redshifts (\cite[Hakopian \& Balayan 2002.2]{HakopianBalayan02.2}, \cite[Hakopian 2002]{Hakopian02}) has been undertaken for the separate samples. The analysis of how different emission lines work in the spectral ranges obtained on photographic plates shows, that a sample can be complete within 250 Mpc (H\(=\)75kms$^{-1}$Mpc$^{-1}$) (\cite[Hakopian 2002]{Hakopian02}). In a sphere of this radius 50\% of the galaxies of nuclear activity of the basic sample (nearly all of the galaxies of type Sy2 and LINER and 6 out of 18 galaxies of type Sy1) are located and 80\% of all galaxies of starforming activity, among which almost all of those in nebular phase, viz. 53 of 55. As currently only two subclasses for SfG galaxies are provided by our scheme, it is easy to get, that among SfGs, of apparent magnitude up to 19-19.5 and of redshift z \textless\ 0.06, on an average, every fifth one is in the more active, nebular phase. In this statistics it must be accounted that an 
effectivity of selection of different types of galaxies in the survey is not the same. In particular, corresponding evaluations for the uv-excess, as a selection criterium, which works in the range up to 18,5 of apparent magnitude, was most effective in the selection of Sy1 type galaxies as well as for galaxies SfGneb, most of which are known as blue compact galaxies, BCG. The latter ones can be seen as the best result by productivity among SBS galaxies, since the second selection criterium used in SBS, i.e. emission lines, which works up to the surveys' limiting, 19,5 magnitude, is also mostly effective for their selection.\\   

\textbf{3.\emph{Current Stage.}}\\ 
In the current stage detailed investigation of separate objects of the basic sample is being carried out, the basis of which are the results of the panoramic spectroscopy with the coverage of spectral range, which includes the recombination  H$\alpha$ line of hydrogen and at least its nearest forbidden doublets of the nitrogen [NII]$\lambda$$\lambda$6548,6583 and sulfur [SII]$\lambda$$\lambda$6717,6731 lines. The observations were started from the most interesting by morphology galaxies with the usage of multipupil spectrographs MPFS and VAGR, with the 6-m telescope of SAO and the 2.6m telescope of BAO, respectively (e.g.,\cite[Hakopian et al. 2006]{Hakopian_etal06}, \cite[Hakopian et al. 2012]{Hakopian_etal12}).
The development of more detailed scheme of classification of starforming galaxies is within the goals connected to further explorations of our basic sample.\\

\end{document}